\pdfoutput=1
\documentclass[11pt]{article}
\usepackage{jheppub}
\usepackage{color}
\usepackage{amsfonts}
\usepackage{amsmath,amssymb,amsthm}
\usepackage{mathrsfs}
\usepackage{graphicx}
\usepackage[utf8]{inputenc}

\newcommand{\spheno}{\texttt{SPheno} 3.3.8 \cite{Porod:2003um,Porod:2011nf}}
\newcommand{\vevacious}{\texttt{Vevacious} 1.2.02 \cite{Camargo-Molina:2013qva}}
\newcommand{\cosmotransitions}{\texttt{CosmoTransitions} \cite{Wainwright:2011kj}}
\newcommand{\homps}{\texttt{HOM4PS2} \cite{hom4ps:2008}}
\newcommand{\flavio}{\texttt{flavio} \cite{david_straub_2017_580247}}
\newcommand{\sarah}{\texttt{SARAH} 4.11.0 \cite{Staub:2008uz,Staub:2013tta,Porod:2014xia}}
\newcommand{\flavorkit}{\texttt{FlavorKit} \cite{Porod:2014xia}}
\newcommand{\julia}{\texttt{Julia} 0.5.2 \cite{2012arXiv1209.5145B, doi:10.1137/141000671}}
\newcommand{\matplotlib}{\texttt{matplotlib} 2.0.1 \cite{Hunter:2007, michael_droettboom_2017_573577}}
\newcommand{\sushi}{\texttt{SusHi} 1.6.1 \cite{Harlander:2012pb,Harlander:2016hcx,Harlander:2002wh,Harlander:2003ai,Aglietti:2004nj,Bonciani:2010ms,Degrassi:2010eu,Degrassi:2011vq,Degrassi:2012vt,Harlander:2005rq,Chetyrkin:2000yt}}
\newcommand{\lhapdf}{\texttt{LHAPDF} 6.1.6 \cite{Gomes:2013qza}}
\newcommand{\pdf}{\texttt{MMHT 2014}~\cite{Harland-Lang:2014zoa}}

\author[a, b]{Anibal D.~Medina}
\emailAdd{anibal.medina@fisica.unlp.edu.ar}
\author[c]{Michael A.~Schmidt}
\emailAdd{michael.schmidt@sydney.edu.au}

\affiliation[a]{Institut de Physique Th\'eorique, Universit\'e Paris Saclay, CNRS, CEA, F-91191 Gif-sur-Yvette, France}
\affiliation[b]{IFLP, CONICET-Departamento de F\'isica,  Universidad Nacional de La Plata, C.C 67, 1900 La Plata, Argentina}
\affiliation[c]{ARC Centre of Excellence for Particle Physics at the Terascale, School of Physics, The University of Sydney, Physics Road, NSW 2006, Australia} 

\title{\mathversion{bold}Enlarging Regions of the MSSM Parameter Space for Large $\tan\beta$ via SUSY Decays of the Heavy Higgs Bosons}

\abstract{In the Minimal Supersymmetric Standard Model (MSSM) 
searches for the heaviest CP-even and CP-odd Higgs $H$, $A$ to tau-lepton pairs severely constrain the parameter region for large values of $\tan\beta$ and 
light Higgs bosons $H$, $A$. We demonstrate how the experimental constraint can be avoided by new decays to light third-generation sfermions,
whose left-right couplings to $H$ can be maximised in regions of large trilinear couplings $A_{b}$, $A_{\tau}$ for sbottoms and staus, or large supersymmetric (SUSY) Higgs mass $\mu$ for stops.
Due to the
$\tan\beta$-enhancement in the production cross-sections via gluon-fusion and
in association with bottom-quark pairs for $H$ and $A$, we find that down-type
sfermions, in particular, sbottoms perform a better job in allowing more
parameter space than up-type sfermions such as stops, which require much larger
values of $\mu$ to compensate for $\tan\beta$. Vacuum stability 
as well as flavour observables constraints
and direct searches for SUSY particles are imposed. We also associate the
lightest CP-even Higgs with the observed 125 GeV SM-like Higgs and impose the
experimental constraints from the LHC.}

\begin{document}

	\maketitle


\section{Introduction}
More than one Higgs doublet is expected in many theories beyond the Standard
Model (SM). The most economical and well-studied supersymmetric extension of
the SM, the Minimal Supersymmetric Standard Model (MSSM),  contains a type-II two Higgs doublet system due to holomorphicity in its electroweak symmetry breaking sector (EWSB). If CP is a good symmetry of the Higgs sector, the scalar Higgs
spectrum consists of two CP even Higgs bosons $h$ and $H$, one CP odd Higgs $A$ and
charged Higgs pair $H^{\pm}$; the lightest CP-even Higgs $h$ is most easily
identified with the 125 GeV SM-like Higgs resonance discovered at the LHC.  On
the other hand, heavier neutral Higgs bosons are being searched for at the LHC via
their decay into a pair of tau-leptons and strong constraints are put on the
allowed masses as a function of the ratio of the Higgs doublet vacuum
expectation values (vev) $\tan\beta\equiv \langle H_u \rangle/\langle H_d
\rangle = v_u/v_d$. In particular, the latest CMS~\cite{CMS:2016rjp} and ATLAS~\cite{ATLAS:2016fpj} results show that
if only decays to SM fermions and gauge bosons are considered, then
$m_{H}\approx m_{A}> 500$ GeV for $\tan\beta \gtrsim 20$, ruling out regions of
large $\tan\beta$ and moderate $m_A$. These regions however are very appealing
since for $\tan\beta\gg 1$ there is an apparent unification of Yukawa couplings
$y_t\approx y_b\approx y_{\tau}$ and also a somewhat light Higgs sector has
better chances of being probed at the LHC. 
Furthermore, for $\tan\beta\gg 1$, the off-diagonal mass mixing between the SM Higgs and the non-standard Higgs~\footnote{ In the Higgs basis $h$ is the SM Higgs whose vacuum expectation value (vev) $\langle h \rangle=v=246$ GeV, whereas $H$ is the
non-standard Higgs that has vanishing vev $\langle H \rangle=0$.} is suppressed as
$\sin 2\beta\sim 1/\tan\beta$, which is very easy to see in the so-called "Higgs basis"~\cite{Carena:2013ooa}.
This conclusion holds even with
the inclusion of finite radiative corrections which are important to bring the
lighter Higgs mass eigenstate to 125 GeV. 

In this work we show that
these constrained regions can be consistent with collider searches if there are
additional decays for the heavy Higgs bosons which suppress the branching ratios of
$H$ and $A$ to tau-leptons, Br$(A,H\to\tau\bar{\tau})$.  For that purpose we
consider the possibility of having additional decays into pairs of sbottoms,
stops and staus respectively. This has also been suggested in Ref.~\cite{Bartl:1996vh,Carena:2013iba,Barman:2016jov} and studied in detail for electroweakinos in Ref.~\cite{Djouadi:2015jea,Barman:2016kgt}, where in the latter it was shown that SUSY decays into electroweakinos can be relevant for values of $5\lesssim\tan\beta\lesssim 20$. We go beyond these studies by analysing the possible consequences on the
destabilisation of the electroweak vacuum and flavour
violating contributions, which impose an important
constraint on the possible branching ratio to
sfermions. We take into account the latest constraints
on direct production of
these SUSY particles. In particular, we exploit the left-right (LR) coupling of
the heavy Higgs bosons to a pair of down-type sfermions which has a term
proportional to $A_{f} \tan\beta$, that allows firstly to
overcome the $\tan\beta$ enhancement of the usual dominant bottom-quark
contribution to the total decay width  and then to even possibly dominate the
total decay for a sufficiently large value of the trilinear coupling
$A_{f}$. In the case of stops we find it necessary to consider large
values of the Higgs SUSY conserving mass $\mu$ in order to overcome the
$\tan\beta$ enhancement. 

We perform a numerical study and scan the parameter
space, calculating the production cross-section for $H,A$ via gluon-fusion and
in association with two bottom-quarks with
\sushi\footnote{See
Refs.~\cite{Bartl:1997yd,Eberl:1999he} for earlier
calculations of QCD corrections to the decays of heavy
Higgs bosons to quarks and squarks.}, and the decays
and flavour
observables with \sarah, \spheno, and \flavio. Finally we study possible stability issues with \vevacious, which tend to constrain the maximum allowed values for
$A_{f}$ and $\mu$. We find that indeed it is possible to partially
recover some regions of the $m_{A}$-$\tan\beta$ plane which seem to be
disfavoured by current di-tau searches, enlarging the allowed large $\tan\beta$
regions in the MSSM.

The main theoretical considerations are discussed in Sec.~\ref{sec:theory}. Our results for light sbottoms are presented in Sec.~\ref{sec:sbottoms}, for light staus in Sec.~\ref{sec:staus}, and for light stops in Sec.~\ref{sec:stops}.
Finally, we
conclude in Sec.~\ref{sec:conclusions}.


\section{Analytical Motivation}
\label{sec:theory}
We start with the tree-level coupling and decay rate expressions for the heavy
Higgs bosons to fermions and sfermions. These expressions are well known and can be
found for example in Ref.~\cite{Djouadi:1996pj}. We focus on the case of
down-type fermions for which the couplings and decay rate take the form,
\begin{equation}
\Gamma(\Phi\to d\bar{d})=N_c\frac{G_F M_{\Phi}}{4\sqrt{2} \pi}m^2_{d}g^2_{\Phi \bar{d}d}\beta^p_{d} 
\end{equation}
where $\Phi=H,A$, $N_c$ is the colour factor,  $p=3,1$ for CP-even or odd Higgs bosons, $\beta_{d}=(1-4m^2_d/M^2_{\Phi})^{1/2}$ and
\begin{equation}
g_{H\bar{d}d}=\frac{\cos\alpha}{\cos\beta}, \qquad g_{A\bar{d}d}=\tan\beta
\end{equation}
with $\alpha$ the usual Higgs mixing angle that relates the flavour to the mass
eigenbasis.  In fact, when $\tan\beta\gg 1$,  $\alpha\to \beta-\pi/2$ with the
lightest Higgs CP-even mass eigenstate SM-like, implying that
$g_{H\bar{d}d}\to\tan\beta$, so we see that both couplings are enhanced by
$\tan\beta$. We should mention that couplings to up-type quarks on the other
hand are suppressed in the same limit by $1/\tan\beta$.  From these expressions
one can readily calculate Br$(H,A\to\tau\bar{\tau})$ when only SM-particle
decays are allowed since the total decay width is dominated by decays to
bottom-quarks and find that Br$(H,A\to\tau\bar{\tau})\approx 0.1$,
independently of $\tan\beta$ and $M_{\Phi}$. Thus this branching ratio is
fixed. The dominant production mechanisms for $\Phi$, as mentioned before,  are
gluon fusion and production in association with bottom-quarks, with the latter
dominating the production for very large values of $\tan\beta$. Given that
$g_{\Phi\bar{d}d}$ enters linearly in both production diagrams, we clearly see
that there will be a dependence of the form $\sigma_{H,A}\times
\mathrm{Br}(H,A\to\tau\bar{\tau})\propto \tan^2\beta$, where $\sigma_{H,A}$ represents
both production mechanisms, from which we understand how the constraints for
large values of $\tan\beta$ come about.

The couplings and decay rates for sfermions take the form,
\begin{equation}
\Gamma(\Phi\to\tilde{f}_i\tilde{f}_j)=N_c\frac{G_F}{2\sqrt{2}\pi M_{\Phi}}\lambda^{1/2}_{\tilde{f}_i\tilde{f}_j \Phi} g^2_{\Phi\tilde{f}_i\tilde{f}_j}\label{Eq:sdowndecay}
\end{equation}
with $\tilde{f}_i$, $i=1,2$ the sfermion mass eigenstates and
$\lambda_{\tilde{f}_i\tilde{f}_j \Phi}$ is the well known Kallen
lambda-function which appears in the kinematics of a two-body decay,
\begin{equation}
\lambda_{ijk}=\left(1-\frac{M^2_i}{M^2_k}-\frac{M^2_j}{M^2_k}\right)^2-4\frac{M^2_i M^2_j}{M^4_k}\;.
\end{equation}
 Notice that contrary to the case of decay to fermions which grows with
 $M_{\Phi}$, decays to sfermions are suppress by $1/M_{\Phi}$. The couplings
 $g_{\Phi\tilde{f}_i\tilde{f}_j}$ are combinations of chiral-couplings,
\begin{equation}
g_{\Phi\tilde{f}_i\tilde{f}_j}=\sum_{\alpha,\beta=L,R}T_{ij\alpha\beta}g_{\Phi \tilde{f}_{\alpha}\tilde{f}_{\beta}}\;.
\end{equation}
The couplings with the same chirality have terms proportional to SM fermions or
gauge boson masses and thus are not efficient in enhancing these couplings.
Interestingly, the mixed-chirality couplings take the form,
\begin{eqnarray}
g_{A \tilde{d}_{L}\tilde{d}_{R}} &=&-\frac{1}{2}m_d\left[\mu +A_{d}\tan\beta\right],\qquad \;\;\;\;g_{H \tilde{d}_{L}\tilde{d}_{R}}=-\frac{1}{2}m_d\left[\frac{\sin\alpha}{\cos\beta}\mu +A_{d}\frac{\cos\alpha}{\cos\beta}\right]\nonumber\\
g_{A \tilde{u}_{L}\tilde{u}_{R}} &=&-\frac{1}{2}m_u\left[\mu -\frac{1}{\tan\beta}A_{u}\right],\qquad g_{H \tilde{u}_{L}\tilde{u}_{R}}=-\frac{1}{2}m_u\left[\frac{\cos\alpha}{\sin\beta}\mu +A_{u}\frac{\sin\alpha}{\sin\beta}\right]\label{Eq:chiralcoupl}
\end{eqnarray}
which depend on the SUSY breaking trilinear couplings $A_f$ and SUSY conserving mass $\mu$. Thus in
the large $\tan\beta$ regions we see that there will be terms enhanced by
$\tan\beta$ proportional to $A_{d}$ which can be used to increase the couplings
to down-type sfermions. In the case of couplings to up-type sfermions, there is
only at most  a term independent of $\tan\beta$ growing with $\mu$ which can be
used to increase the coupling~\footnote{The trilinear interaction for up-type sfermions is suppressed by $1/\tan\beta$ as shown in Eq.(\ref{Eq:chiralcoupl}).}. The factor $T_{ij\alpha\beta}$ takes into
account the chiral mixing in the mass basis and in order to maximise the
couplings we should be close to maximal mixing
$\sin\theta_f\approx\cos\theta_f\approx 1/\sqrt{2}$, with $\theta_f$ the mixing
angle.  We take also into account important loop-level contributions which
modify the relation between down-type Yukawas and running masses,
\begin{equation}
y_{b}=\frac{m_b}{v\cos\beta(1+\Delta_b)},\qquad y_{\tau}=\frac{m_{\tau}}{v\cos\beta(1+\Delta_{\tau})}
\label{Eq:ybytau}
\end{equation}
where $\Delta_{b}$ is dominated by sbottom-gluino and stop-chargino loop, whereas $\Delta_{\tau}$ is dominated by stau-neutralino and sneutrino-chargino loop, and both can be sizeable in the large $\tan\beta$ regime.

Given that we want the contribution from the L-R coupling to be maximal in
order to enhance the decays into SUSY particles, we must choose the soft
breaking masses to be roughly of the same order in particular for the sbottom
and stau sectors due to their smaller Yukawa couplings. This implies that we
expect $m_{\tilde{b}_2}\gtrsim m_{\tilde{b}_1}$ and $m_{\tilde{\tau}_2} \gtrsim
m_{\tilde{\tau}_1}$, which we find in our numerical studies. For stops the
story is different given their important loop contribution to the effective
Higgs potential which pushes the lightest Higgs mass to 125 GeV. In this case
 one must choose one of the soft breaking masses ($m_{U_3}$ or $m_{Q_3}$) to
be of the same magnitude as $A_t\simeq 2$ TeV, pushing the heavier stop in the
few TeV region. However, since we want to have the heavy Higgs bosons decay to stops
in the first place, we need the lighter stop to remain light enough to
kinematically allow for such decays.  Thus in this case the spectrum is more
split ($m_{\tilde{t}_1}\ll m_{\tilde{t}_2}$ and $m_{\tilde{t}_1}\lesssim
m_{\Phi}/2$) than for sbottoms and staus and though the mixing is not maximal
we can still have stops contributing to the total decay width enough to
suppress Br$(\Phi\to\tau\bar{\tau})$.

Large values of $A_{b}$, $A_{\tau}$ and $\mu$ are constrained by colour and electromagnetic charge breaking since they provide cubic terms in the scalar potential that tend to
destabilise the neutral electroweak symmetry breaking vacuum~\cite{Carena:2012mw}. We
will see in the next section that this puts strong constraints on the allowed
values for $A_{b}$, $A_{\tau}$ and $\mu$. There are flavour violating processes
which are enhanced at large $\tan\beta$, in particular B-meson
decays (See e.g.~\cite{Altmannshofer:2012ks}). We will also comment on this in the next sections.
We perform a numerical scan for each of the three discussed possibilities. 

\section{Light Sbottoms}
\label{sec:sbottoms}
We describe the parameter space and the codes used in the numerical scan for light sbottom quarks in the next section, before discussing our results in Sec.~\ref{sec:sbottomsResult}.

\subsection{Numerical Scan}
In order to study the feasibility to enlarge regions of large values of $\tan\beta$ currently constrained by $H,A\to \tau\bar\tau$ searches, we 
do a numerical simulation of the productions of $H$ and $A$ via gluon fusion
and in association with bottom-quark pairs using \sushi, a Fortran code which
can calculate these production cross sections in the MSSM.  In the case of
gluon fusion, it takes into account  $NLO$ QCD contributions from the third
family of quarks and squarks, $N^{3}LO$ corrections due to top-quarks,
approximate $NNLO$ corrections due to top squarks and electroweak effects.
Very much relevant for large values of $\tan\beta$ for the down-type sector and
it particular for the third family Yukawa couplings, it resums higher order $\tan\beta$-enhanced sbottom contributions.
The supersymmetric particle spectrum, as well as cross-sections  and decays for
SUSY particles, are calculated using \sarah\ and \spheno, in particular the \texttt{SPheno} version generated from the MSSM model file in \texttt{SARAH}. 
We subsequently calculate flavour observables with \flavio, which takes the Wilson coefficients calculated by \flavorkit\ as input, and the Higgs production cross sections at the LHC with \sushi\ for both CP even Higgs bosons using the \pdf\ parton distribution functions set via \lhapdf.
Stability  of the electroweak vacuum and
possible charge$/$colour breaking minima are investigated using \vevacious, which relies on \cosmotransitions\ and \homps. 
Due to the lack of SUSY signals so far at the LHC, we decide to consider a
$natural$ spectrum, pushing 1\textsuperscript{st} and 2\textsuperscript{nd}-generation sparticles, as
well as gluinos and Winos in the multi-TeV range:
\begin{equation}
	m_{\tilde e_j}=m_{\tilde L_j}=m_{\tilde u_i}=m_{\tilde d_i} = m_{\tilde Q_i} = M_2=M_3 = 2.2\; \mathrm{TeV} 
\end{equation}
with vanishing $A$-terms.
For 3\textsuperscript{rd}-generation
sparticles, depending on how we want to suppress the branching ratio Br$(H,A\to\tau\bar\tau)$, we
keep either sbottoms, staus or stops light~\footnote{In order to obtain large
enough radiative corrections to increase the light Higgs mass to $\sim$ 125
GeV, one tends to need large values of $A_t$ which can lead to a light stop in
the spectrum.} to allow for heavy Higgs SUSY decays to be kinematically accessible.
Since $|\mu |, M_1\ll M_2, M_3, m_{\tilde{f}_{1,2}}$, the other possible light
sparticles in the spectrum are the first three lighter neutralinos
$\tilde{\chi}^0_1$, $\tilde{\chi}^0_2$, $\tilde{\chi}^0_3$ and the light
chargino $\tilde{\chi}^{\pm}_1$.
In the scan with light sbottoms, we fixed
\begin{align}
	M_1& = 200\; \mathrm{GeV}&
	m_{\tilde u_3} &= 2845\; \mathrm{GeV}\;.
\end{align}
and varied the remaining parameters, $\tan\beta$, $\mu$, $B_\mu$, $m_{\tilde Q_3}$,  $m_{\tilde d_3}$, and $A_t$
\begin{align}
	\tan\beta & \in [25,60] &
	m_{\tilde Q_3} & \in [300,800]\; \mathrm{GeV} &
	m_{\tilde d_3} & \in [300,800]\; \mathrm{GeV} \\\nonumber
	\mu & \in \pm  [200,400]\; \mathrm{GeV} &
	m_A(\mathrm{tree}) & \in [500,1600]\; \mathrm{GeV} &
	A_t  & = \pm m_{\tilde u_3}\;.
\end{align}
For all points with $m_{\tilde b_1}\geq 300$ GeV, which are close to the
experimental exclusion limit of the $H\to\tau\bar\tau$ searches and for which the decay
$H\to \tilde b \tilde b^*$ is kinematically accessible, we increased $|A_b|$, both for positive and negative
$A_b$,  and used a fixed-point iteration to determine the largest possible value, for which the electroweak
vacuum is either stable or sufficiently long-lived on cosmological scales. Finally, we increased $M_1$ to enlarge the parameter space by further suppressing the limits from the direct sbottom pair production searches.

We impose that the lightest Higgs particle in the spectrum, which is associated with the scalar resonance discovered at
the LHC, satisfies the measurement of the Higgs mass $125\pm 3$ GeV taking the theory error into account and the latest signal strengths measurements by ATLAS and CMS at the $2\sigma$ level (Tab.~16 in Ref.~\cite{Khachatryan:2016vau}), in the different relevant channels: $b\bar{b}$, $WW^{*}$, $ZZ^{*}$, $\tau\bar\tau$
and $\gamma\gamma$. 
We discarded all data points, which have a sbottom quark mass below $300$ GeV
to satisfy mono-jet searches at 3.2 fb$^{-1}$\cite{Aad:2015pfx} and directly
use the latest 13 TeV CMS direct sbottom\cite{CMS:2017uwv} and
stop\cite{CMS:2017qjo} pair production searches with a luminosity
$\mathcal{L}=36.1 \mathrm{fb}^{-1}$ by imposing the limit extracted from the
provided root files, where we use the QCD squark pair production cross section reported in Ref.~\cite{Borschensky:2014cia}.

For the main object
of our study, the heavy Higgs bosons $H,A$, we require that both the productions in
association with bottom-quarks and via gluon fusion, with subsequent decay into
tau pairs,  $\sigma_{bbH}\times \mathrm{Br}(H\to\tau\bar\tau)$ and $\sigma_{ggH}\times
\mathrm{Br}(H\to\tau\bar\tau)$, satisfy the bounds from both ATLAS~\cite{ATLAS:2016fpj} and CMS~\cite{CMS:2016rjp} studies at 13 TeV
and 13.3 fb$^{-1}$ and 12.9 fb$^{-1}$, respectively, though due to the large values of $\tan\beta$ we are
interested in, the production in association with bottom-quark pairs places stronger
constraints.

We make a few comments with respect to flavour observables and constraints. We
are able to satisfy all flavour observable constraints ($B_s \to
\mu^{+}\mu^{-}$~\cite{Aaij:2017vad}, $B\to\tau\nu$~\cite{Olive:2016xmw}, etc)
at the 2$\sigma$ level, except for $B\to
X_s\gamma$~\cite{Olive:2016xmw}, for which the stop-chargino loop contribution can be significant,
whereas the charged Higgs contributions seems to be subdominant. Within the
Minimal Flavour violation (MFV) paradigm,  a study done in
Ref.~\cite{Altmannshofer:2012ks} shows that for $A_t>0$,  $\mu\gtrsim 800$ GeV
or $M_{Q_3}\gtrsim 1.3$ TeV are necessary to satisfy the latest measurements.
For $A_t<0$, constraints are much stronger and always require $M_{Q_3}\gtrsim
1.5$ TeV. Since we want to have a light enough sbottom for the heavy Higgs bosons to
decay, this implies that the only possibility would be to have $A_t>0$ and
$\mu\gtrsim 800$ GeV, which would not affect the main conclusions of this
work. However, recall that this is all within the MFV paradigm. Beyond the MFV paradigm, there are new ways to suppress the contribution of the stop-chargino loop, in particular possible
additional diagrams involving gluinos and sbottom-strange mixing, which may be
able to cancel the chargino-stop contributions~\cite{Altmannshofer:2013foa,Arana-Catania:2013pia}. Thus we do not impose in our results the constraint from $B\to X_s \gamma$ due to the caveats just discussed.

\subsection{Results}
\label{sec:sbottomsResult}
\begin{figure}[tbp]
 \begin{minipage}[t]{0.5\textwidth}
  \centering
  \includegraphics[width=\textwidth]{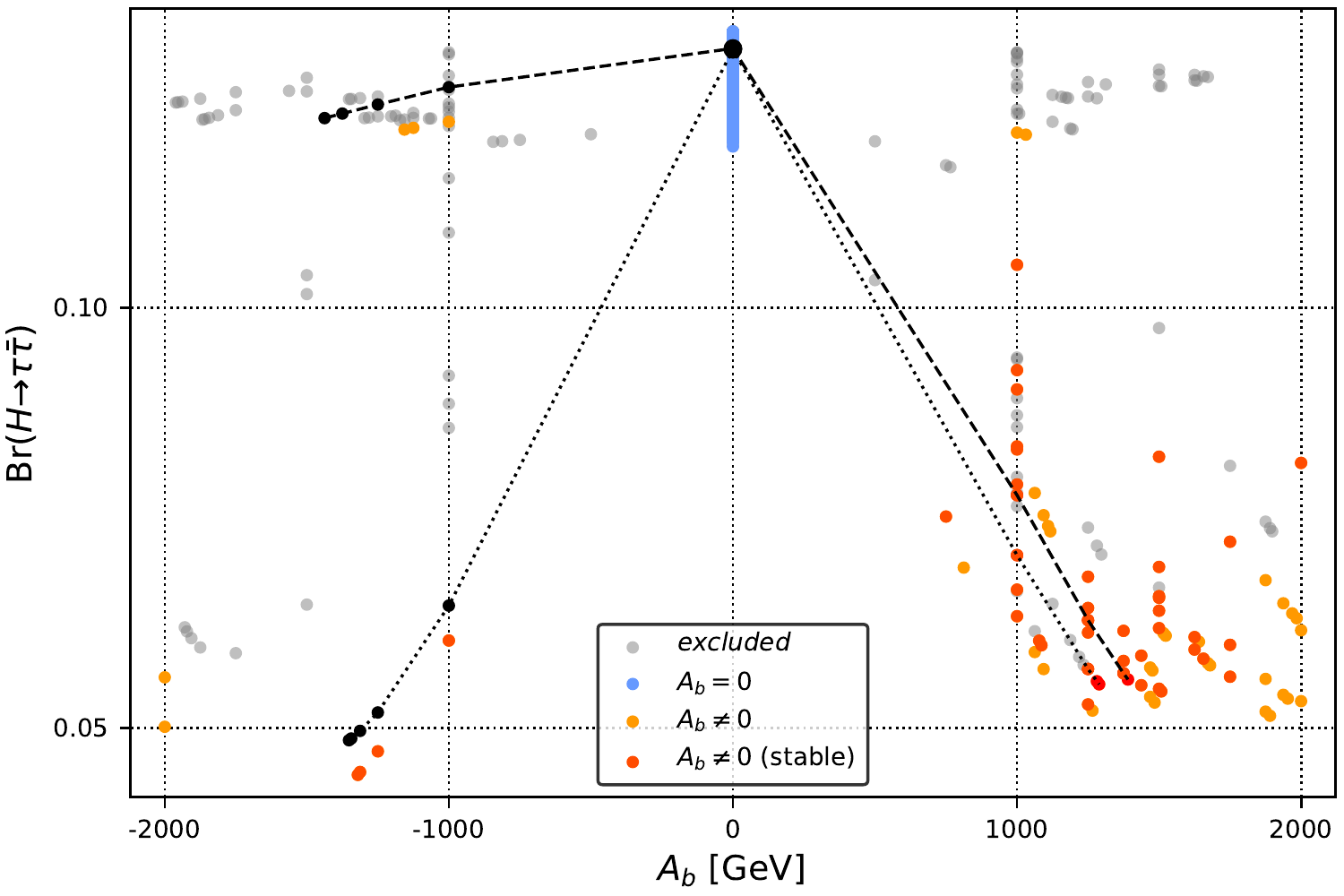}
 \end{minipage}
 \begin{minipage}[b]{0.5\textwidth}
  \centering
  \includegraphics[width=\textwidth]{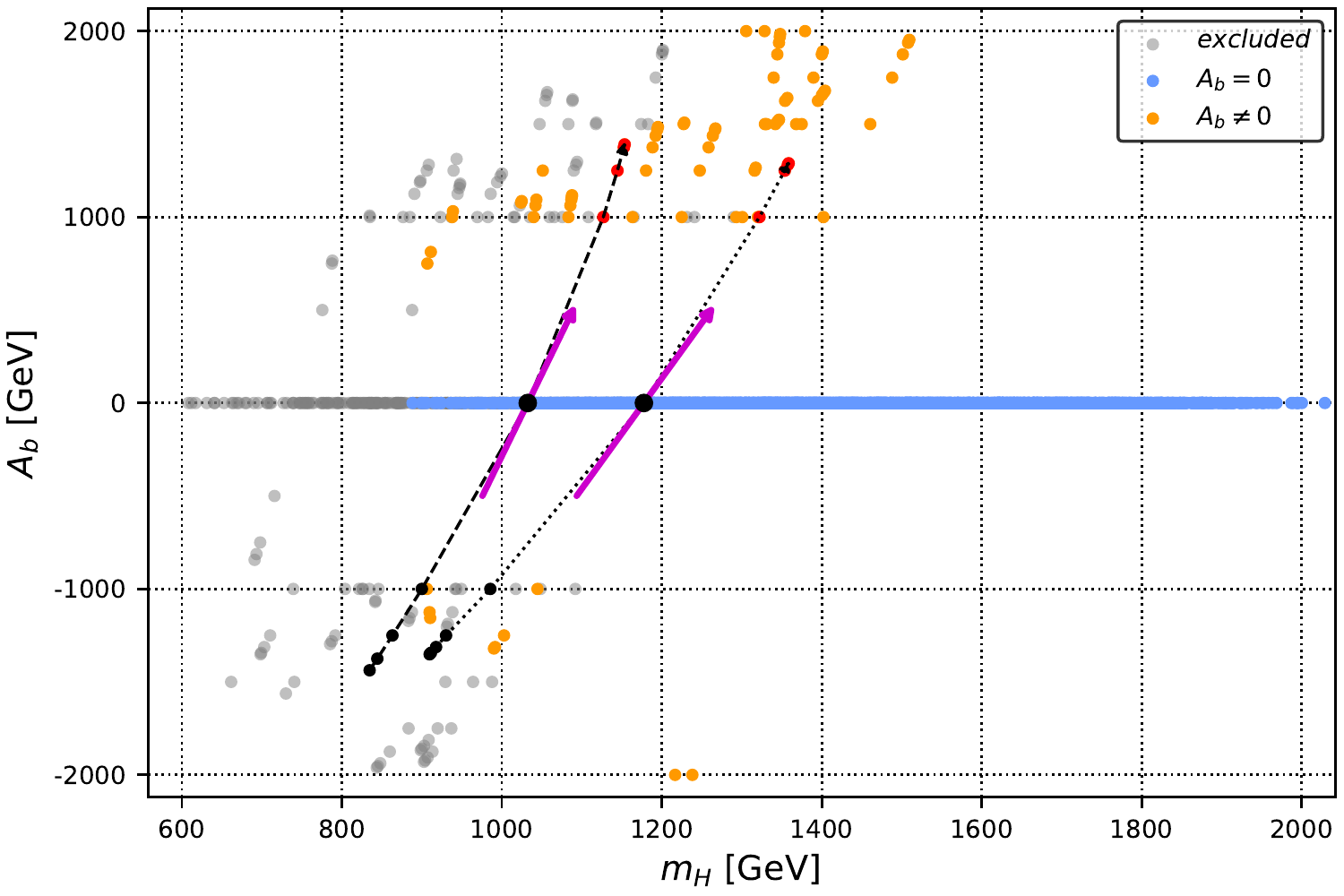}
 \end{minipage}
 \caption{Br$(H\to\tau\bar\tau)$ vs $A_b$ $[{\rm GeV}]$ and $A_b$ $[{\rm
 GeV}]$  vs $m_H$  $[{\rm GeV}]$. Grey points are excluded by $A,H \to
 \tau\bar\tau$ searches, blue points have $A_b=0$ such that decays into sbottoms
 are negligible, orange  and red points have $A_b \neq 0$ and thus a
 non-negligible decay into sbottoms. All orange points have either a fully
 stable or metastable electroweak vacuum. In the left-hand figure, we
 distinguish between the stable and metastable vacuum. Dark orange points in
 the left-hand figure feature a fully stable electroweak vacuum and lighter
 orange points have a metastable electroweak vacuum. Black and red
 points on top of the dashed and dotted black lines are two particular examples
 where we only vary $A_b$ while keeping all other parameters fixed, in order to
 show how we move in the particular planes shown. The purple arrow indicates the direction of increasing $A_b$. Direct sbottom searches as
 well as light Higgs bound constraints are satisfied by all non-excluded points.
 }
 \label{fig:Ab-BrTau-mH}
\end{figure}

\begin{figure}[tb]
 \begin{minipage}[t]{0.5\textwidth}
  \centering
  \includegraphics[width=\textwidth]{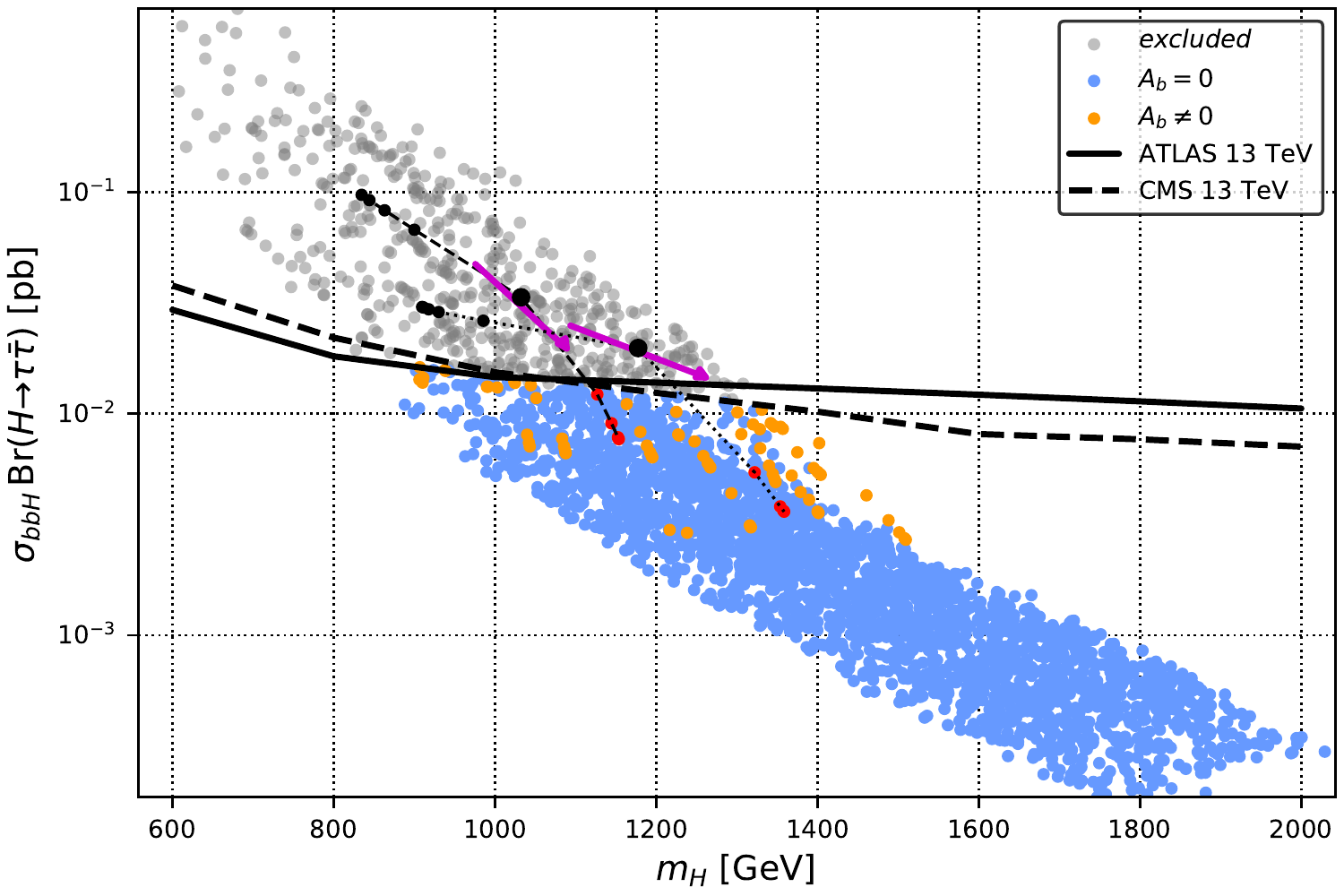}
 \end{minipage}
 \begin{minipage}[b]{0.5\textwidth}
  \centering
  \includegraphics[width=\textwidth]{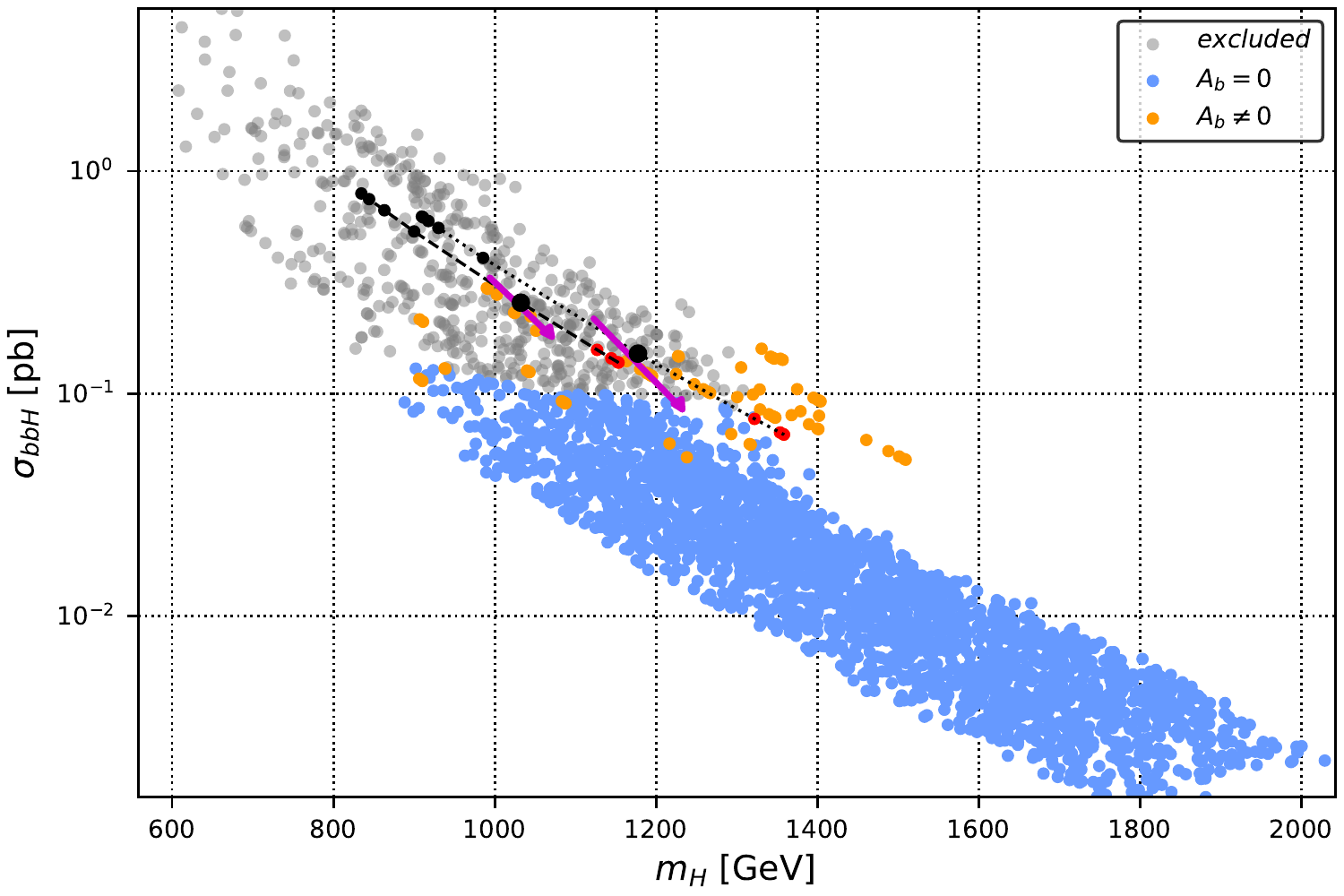}
 \end{minipage}
 \caption{$\sigma_{bbH}\times \mathrm{Br}(H\to\tau\bar\tau)$  $[{\rm pb}]$ vs $m_H$
 $[{\rm GeV}]$ and $\sigma_{bbH}$  $[{\rm pb}]$  vs $m_H$  $[{\rm GeV}]$. Same
 colour coding for points as in Fig.~\ref{fig:Ab-BrTau-mH}. Solid and dashed
 horizontal lines in the figure on the left represent the constraints from the
 latest ATLAS and CMS 13 TeV $A,H \to \tau\bar\tau$ searches at 13.3 fb$^{-1}$,
 respectively. We still show the same two example points where we only vary
 $A_b$. }
 \label{fig:Ab-sigBrBB}
\end{figure}

\begin{figure}[tb]
 \centering
 \includegraphics[height=7cm]{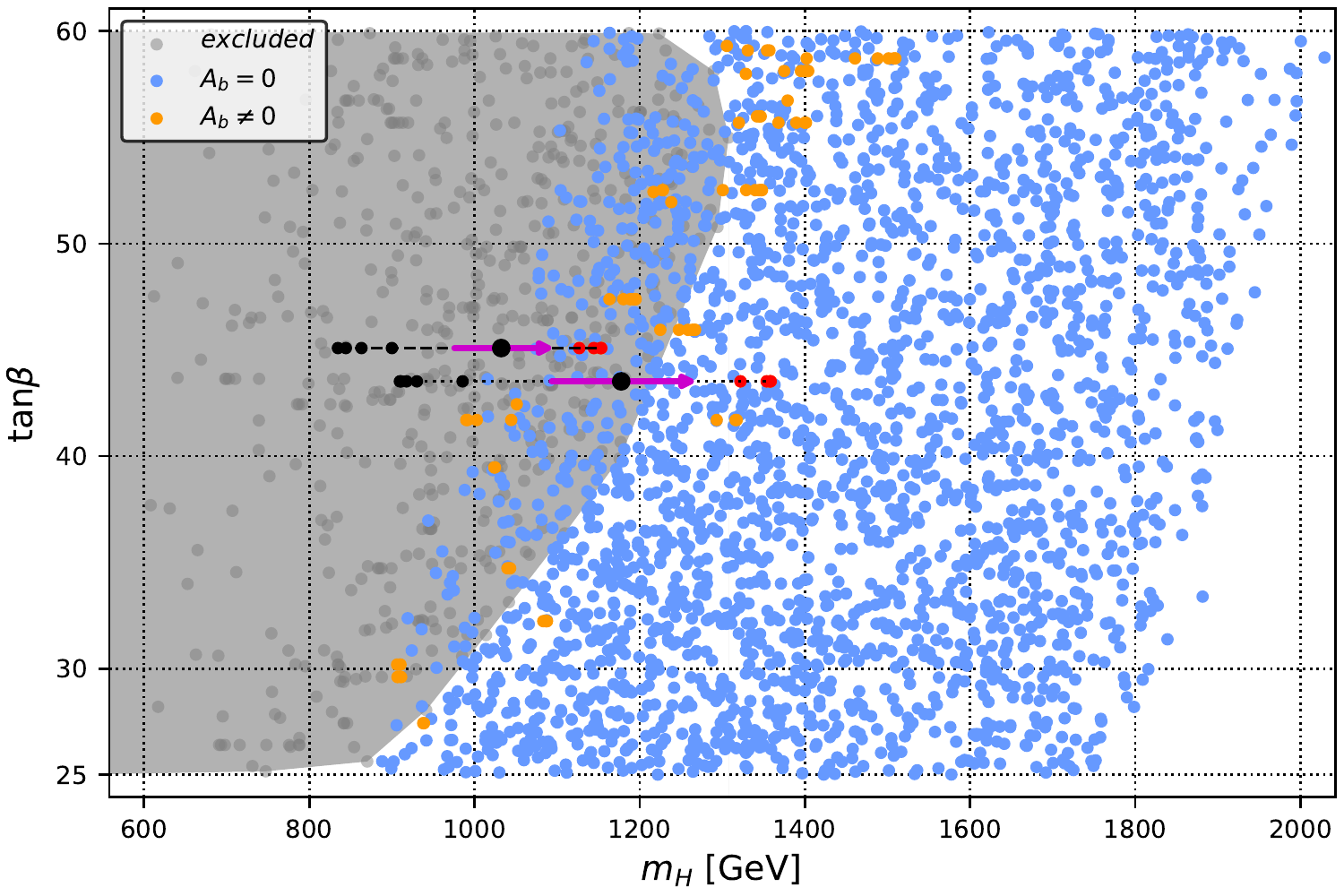}
 \caption{Tan $\beta$ vs $m_H$ $[{\rm GeV}]$. Same colour coding for points as in Fig.~\ref{fig:Ab-BrTau-mH}.  Dark grey region is the envelope of the excluded points. The most interesting points are the orange points that land in the dark grey envelope. We still show the same two example points where we only vary $A_b$.}
 \label{fig:mH-Tanb}
\end{figure}

Having taken in consideration all these constraints, we show our results in
Figs.~\ref{fig:Ab-BrTau-mH}, \ref{fig:Ab-sigBrBB} and \ref{fig:mH-Tanb}. Grey
points are excluded by $A,H \to \tau\bar\tau$ searches, blue points have $A_b=0$
such that decays into sbottoms are negligible, orange  and red points have $A_b
\neq 0$ and thus a non-negligible decay into sbottoms. Orange points feature
either a fully stable or a metastable electroweak vacuum. In the plot on the
left-hand side of Fig.~\ref{fig:Ab-BrTau-mH}, we furthermore indicate data
points with a fully stable electroweak vacuum and $A_b\neq0$ in dark orange. 
The black and red points connected by dashed/dotted lines are two
particular examples where we only vary $A_b$ while keeping all other parameters
fixed, in order to show how we move in the particular planes shown. The big
black dots represent the points with $A_b=0$ and the purple arrow points in the direction of increasing $A_b$.

In Fig.~\ref{fig:Ab-BrTau-mH} on the left, we plot  Br$(H\to\tau\bar\tau)$ vs $A_b$. We immediately see from the red and orange points that as $A_b$ grows in
magnitude, we are able to suppress  the Br$(H\to\tau\bar\tau)$ via the additional
sbottom decays by factors of order a half or slightly smaller. There are
however points which have a large $A_b$ but nonetheless a large
Br$(H\to\tau\bar\tau)$, which implies that these points do not correspond to
maximal mixing between the left and right handed sbottoms. Note as well that we
find both metastable  and fully stable vacua for $|A_b|\lesssim 2$ TeV. In the
two examples shown in this figure  we leave everything fixed except $A_b$ and
one can see that as $A_b$ increases in magnitude one is able to suppress via
decays into sbottoms the Br$(H\to\tau\bar\tau)$. However, for the two  examples the
suppression is not sufficient enough to avoid the LHC constraints from $H,A\to
\tau\bar\tau$ for one the branches ($A_b<0$). Comparing the
location of most of the red and orange points against the grey points, it is
clear that a suppression in Br$(H\to\tau\bar\tau)$ is what allows them to evade the
di-tau constraints. There are, however, some stragglers for which
Br$(H\to\tau\bar\tau)\gtrsim 0.1$ and are able however to evade the constraints.
These points correspond to large $m_H$ such that the constraints from di-taus
ameliorate.  In Fig.~\ref{fig:Ab-BrTau-mH} on the right on the other hand, we
show the influence of $A_b$ on $m_H$. This is clearly seen in the two example
black dashed/dotted lines in this figure, where as we move $A_b$  keeping all other
parameters fixed, we see that $m_H$ can either decrease or increase by several
GeV's, even $\Delta m_H\sim 100$ GeV.  This is coming from the radiative
sbottom corrections to the effective Higgs potential, that as we see for large
values of $\tan\beta$ can be quite relevant~\cite{Carena:1995bx}.  
In the two examples $m_H$ increases with increasing $A_b$, but the opposite behaviour, where $m_H$ decreases with increasing $A_b$, also occurs for some points in the numerical scan. 

This last analysis helps us to partially understand Fig.~\ref{fig:Ab-sigBrBB}. 
In the figure on the right, we show the main constraining production cross
section $\sigma_{bbH}$   as a function of $m_H$. We see that as $m_H$ increases
there is a clear reduction in the cross section as expected. Furthermore, we
see this explicitly in the two examples represented once again by the dashed
black lines. Here we see the effect of $A_b$ shifting $m_H$ and reducing or
increasing the cross section. On the other hand, on the left of
Fig.~\ref{fig:Ab-sigBrBB}, we show $\sigma_{bbH}\times \mathrm{Br}(H\to\tau\bar\tau)$ vs
$m_H$. We also display the constraints from the latest ATLAS~\cite{ATLAS:2016fpj} and CMS~\cite{CMS:2016rjp} $A,H \to \tau\bar\tau$ searches at 13 TeV, respectively, represented by the
solid and dashed nearly horizontal lines in the figure, showing clearly that the grey
points are excluded by these searches. Now we see in the two examples that we
have chosen, that the initial points with $A_b=0$  are right at the border of
exclusion and as we vary $A_b$, we either move into the non-excluded area by
two effects: a decrease in the production cross section due to a larger $m_H$
and a decrease in the Br$(H\to\tau\bar\tau)$ due to di-sbottom decays. Indeed, we
see that the line of the two example becomes steeper as we move in the
non-excluded area. We can however, also move deeper into the excluded  area as
depicted by the two examples,  by a decrease in $m_H$ (which leads to an
increase in the production cross section) and an insufficient suppression of
the branching ratio Br$(H\to\tau\bar\tau)$.

Finally in Fig.~\ref{fig:mH-Tanb}, we show  $\tan\beta$ vs  $m_H$. The grey
envelope area of the excluded points indicates the excluded region. We see something very
interesting happening.  Most of the orange points lie within the grey envelope,
implying that without the additional suppression due to the decays into
sbottom pairs from the heavy Higgs, they would have been ruled out by the current
$H,A\to \tau\bar\tau$ searches. The blue points that lie also within the grey
envelope have a maximal Br$(H\to\tau\bar\tau)\sim 0.1$ as shown in
Fig.~\ref{fig:Ab-BrTau-mH} on the left, but as  seen from
Fig.~\ref{fig:Ab-sigBrBB} on the right, they have a somewhat suppressed
production cross section with respect to the grey points.  This is most likely due to a suppressed coupling from radiative corrections  for these points as
shown in Eq.~(\ref{Eq:ybytau}). This effect has been discussed for decays to stau pairs in Ref.~\cite{Carena:2013iba}.

Focusing on the two example lines, we see that
as we change $A_b$, we move horizontally in the plane $m_H-\tan\beta$ due to
the change in $m_H$ as $A_b$ varies. In both examples we see that starting from
the points with $A_b=0$ which are represented by the slightly larger black
dots, as $m_H$ becomes larger we are able to obtain viable points (red points)
which are within the grey envelope. On the other hand, to the other side of the
big black dot, we move to smaller $m_H$ but are further excluded. These two
behaviours can be understood by looking how in the two examples $m_H$ and
Br$(H\to\tau\bar\tau)$ depend on $A_b$ as shown in Fig.~\ref{fig:Ab-BrTau-mH}. Although Br$(H\to\tau\bar\tau)$ diminishes for increasing $|A_b|$ in both examples, the decrease in Br$(H\to\tau\bar\tau)$ for negative $A_b$ is compensated by the increased production cross section for a lighter $H$. For positive $A_b$, $m_H$ increases with increasing $A_b$ and thus the production cross section is reduced in addition to the suppression of the branching ratio Br$(H\to\tau\bar\tau)$.

\section{Light Staus}
\label{sec:staus}

The numerical scan for light staus is very similar to the one for light sbottoms. We discuss any differences to the scan for sbottoms in the next subsection and our results in Sec.~\ref{sec:stausResult}.
\subsection{Numerical Scan}
We decouple winos, squarks, and the first two generations of sleptons
\begin{equation}
	m_{\tilde Q_j}=m_{\tilde u_j}=m_{\tilde d_j}=m_{\tilde L_i} = m_{\tilde e_i} = M_2=M_3 = 2.2\; \mathrm{TeV} 
\end{equation}
and similarly fix the bino mass and the right-handed stop mass to a large enough loop correction to the Higgs mass 
\begin{align}
	M_1& = 100\; \mathrm{GeV} & 
	m_{\tilde u_3} &= 2845\; \mathrm{GeV}\;.
\end{align}
The other parameters are varied
\begin{align}
	\tan\beta & \in [25,60] &
	m_{\tilde L_3} & \in [150,800]\; \mathrm{GeV} &
	m_{\tilde e_3} & \in [150,800]\; \mathrm{GeV} \\\nonumber
	\mu & \in \pm  [200,400]\; \mathrm{GeV} &
	m_A(\mathrm{tree}) & \in [500,1600]\; \mathrm{GeV} &
	A_t  & = \pm m_{\tilde u_3}\;.
\end{align}
Initially we keep $A_\tau=0$ fixed and in a second step, we increase $|A_\tau|$, and use a fixed-point iteration to determine the largest possible value with a stable or long-lived electroweak vacuum.

Direct stop and sbottom pair production searches are automatically satisfied and we conservatively require $m_{\tilde \tau}\geq 100$ GeV to satisfy the current limits on the $\tilde\tau$ mass~\cite{Olive:2016xmw} and that the lightest supersymmetric particle is a neutralino.
All flavour constraints, and in particular, $B_s \to \mu^{+}\mu^{-}$~\cite{Aaij:2017vad}, $B\to\tau\nu$~\cite{Olive:2016xmw}, $B\to X_s\gamma$~\cite{Olive:2016xmw},
are satisfied at the 2$\sigma$ level. Similarly to the scan with light sbottoms, we impose the Higgs signal strength measurements at $2\sigma$ as well as the Higgs mass measurement.


\subsection{Results}
\label{sec:stausResult}
\begin{figure}[ptb]
 \begin{minipage}[t]{0.5\textwidth}
  \centering
  \includegraphics[width=\textwidth]{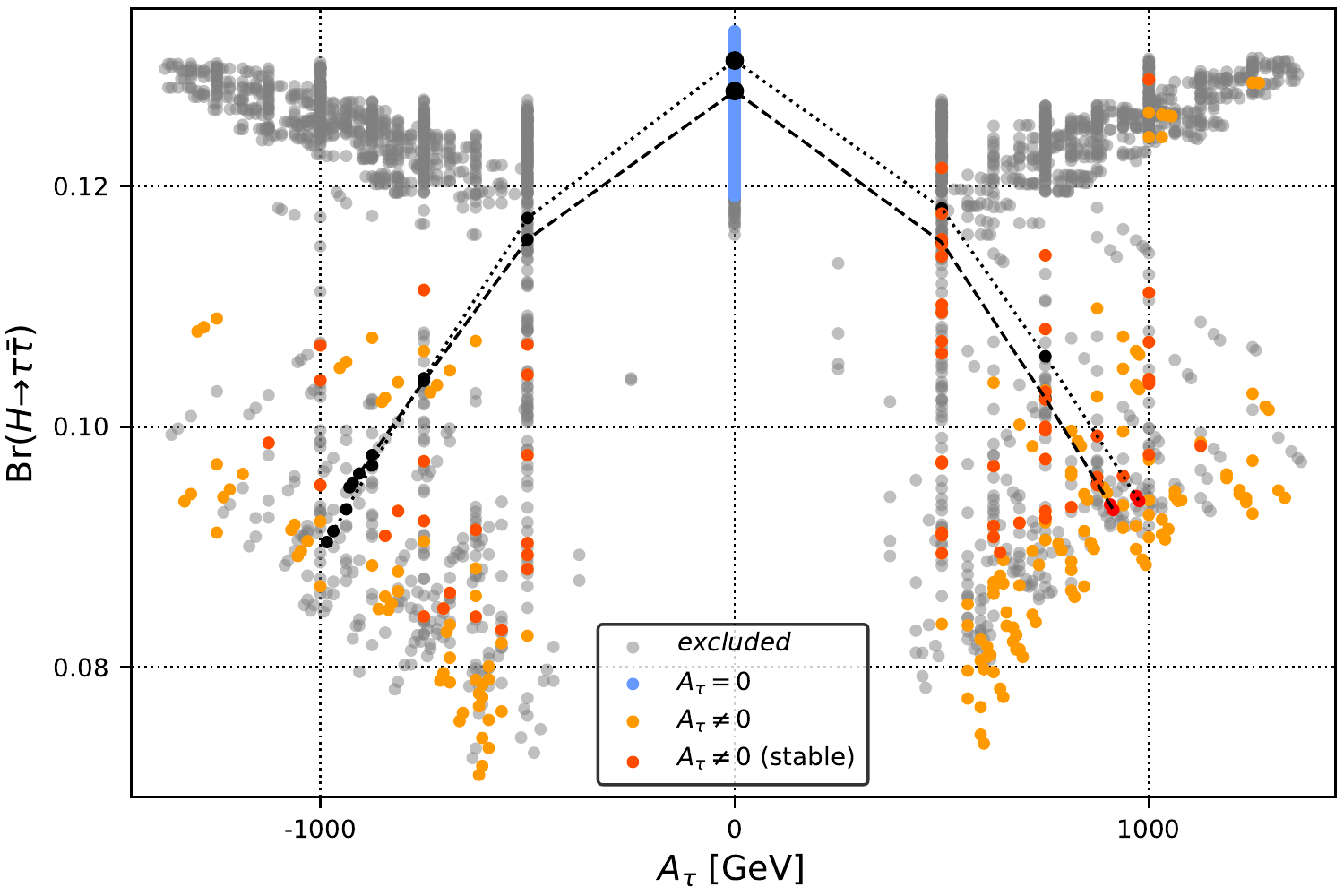}
 \end{minipage}
 \begin{minipage}[b]{0.5\textwidth}
  \centering
  \includegraphics[width=\textwidth]{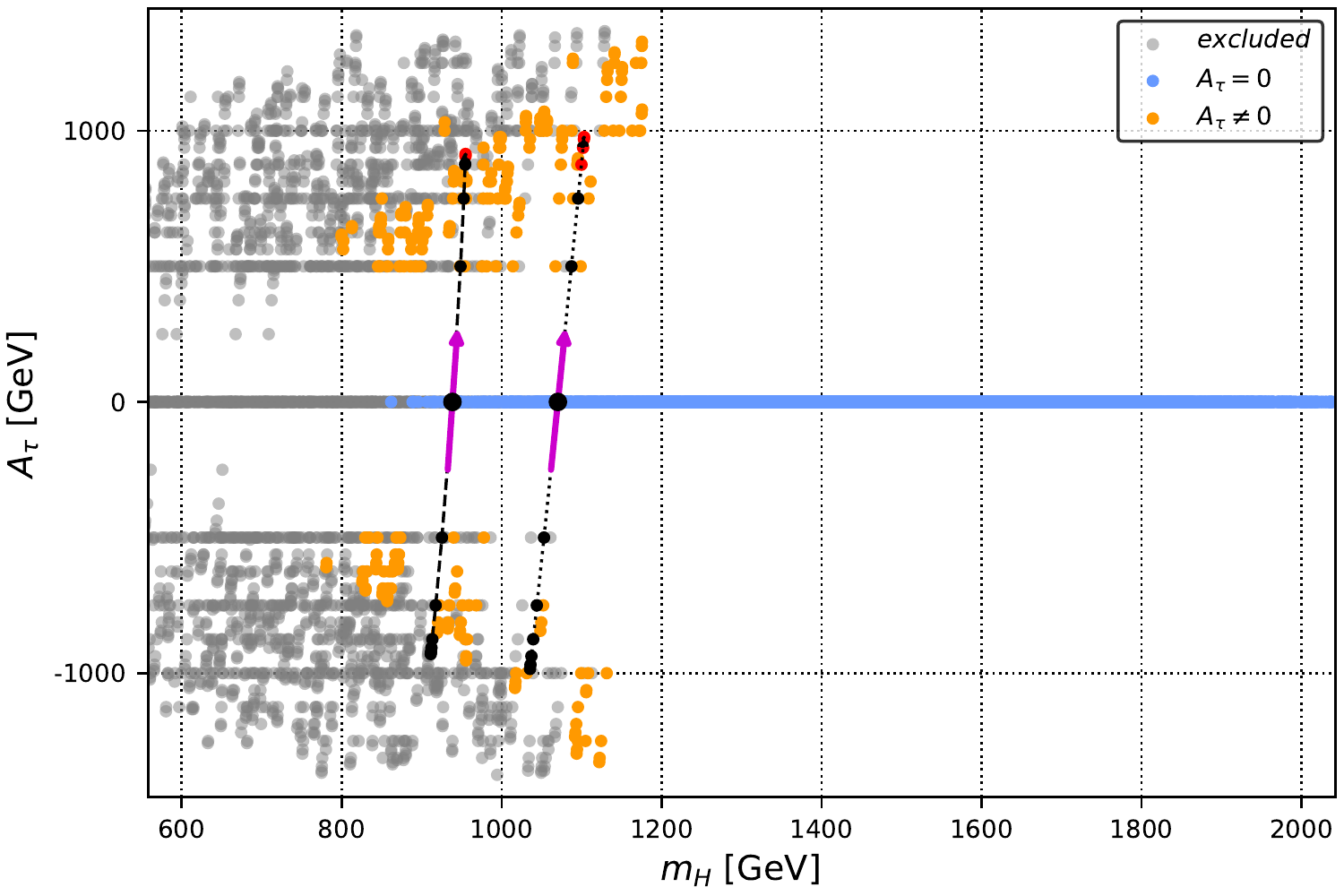}
 \end{minipage}
 \caption{Br$(H\to\tau\bar\tau)$ vs $A_\tau$ $[{\rm GeV}]$ and $A_\tau$ $[{\rm
	 GeV}]$  vs $m_H$  $[{\rm GeV}]$. Same colour coding for points as in Fig.~\ref{fig:Ab-BrTau-mH}. }
 \label{fig:Atau-BrTau-mH}
\end{figure}

\begin{figure}[ptb]
 \begin{minipage}[t]{0.5\textwidth}
  \centering
  \includegraphics[width=\textwidth]{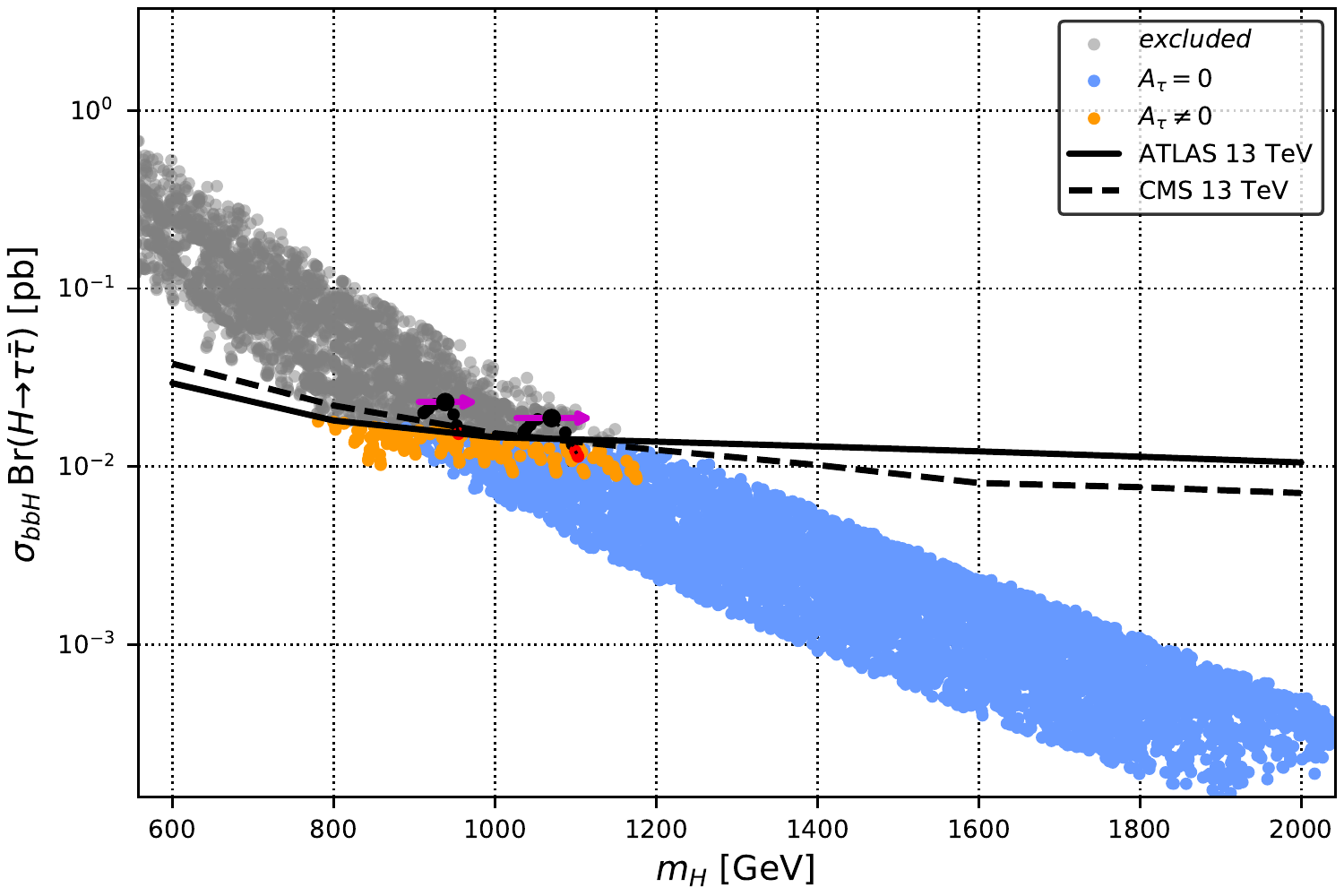}
 \end{minipage}
 \begin{minipage}[b]{0.5\textwidth}
  \centering
  \includegraphics[width=\textwidth]{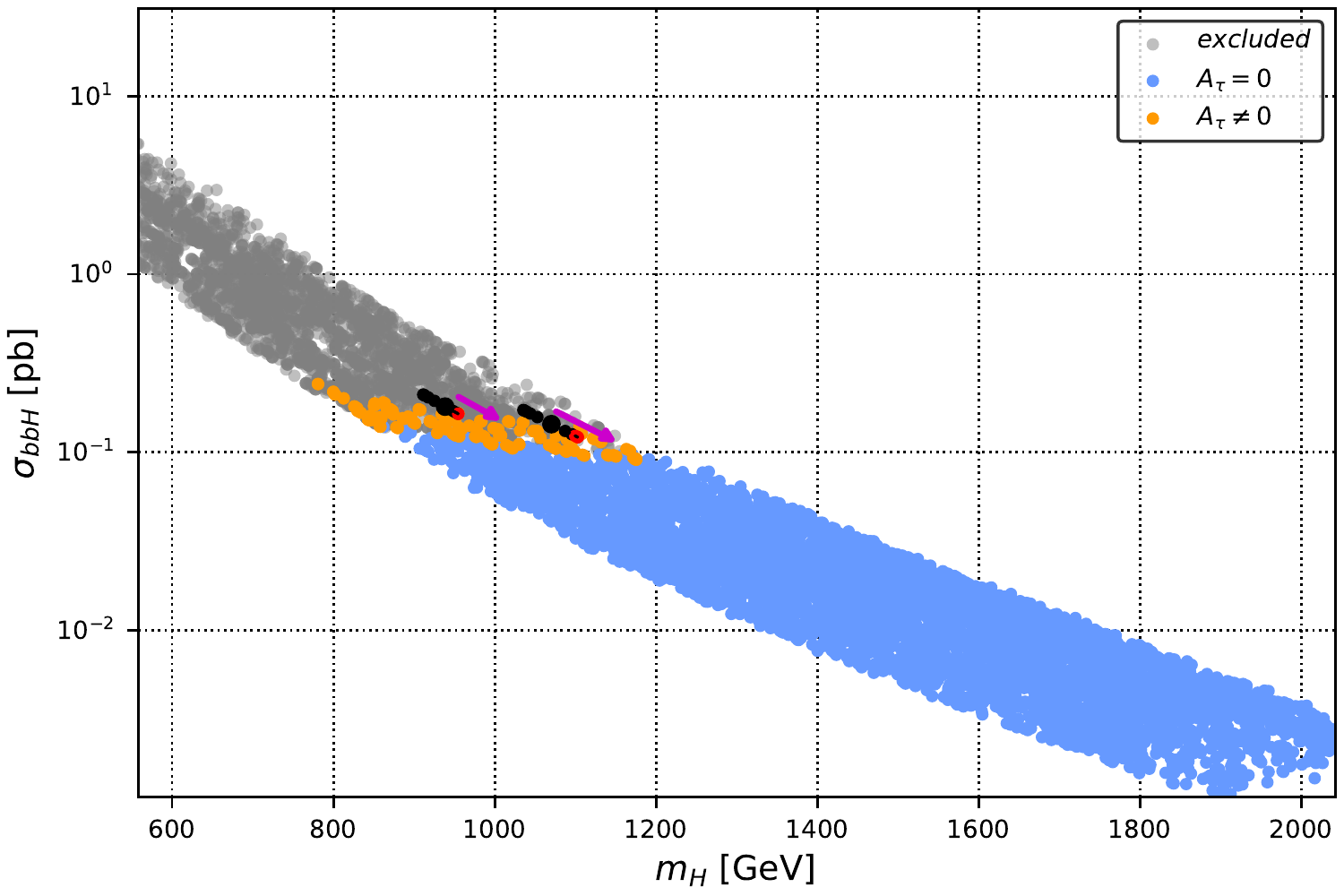}
 \end{minipage}
 \caption{$\sigma_{bbH}\times \mathrm{Br}(H\to\tau\bar\tau)$  $[{\rm pb}]$ vs $m_H$
 $[{\rm GeV}]$ and $\sigma_{bbH}$  $[{\rm pb}]$  vs $m_H$  $[{\rm GeV}]$. Same
 colour coding for points as in Fig.~\ref{fig:Ab-BrTau-mH}. 
 }
 \label{fig:Atau-sigBrBB}
\end{figure}

\begin{figure}[ptb]
 \centering
 \includegraphics[height=7cm]{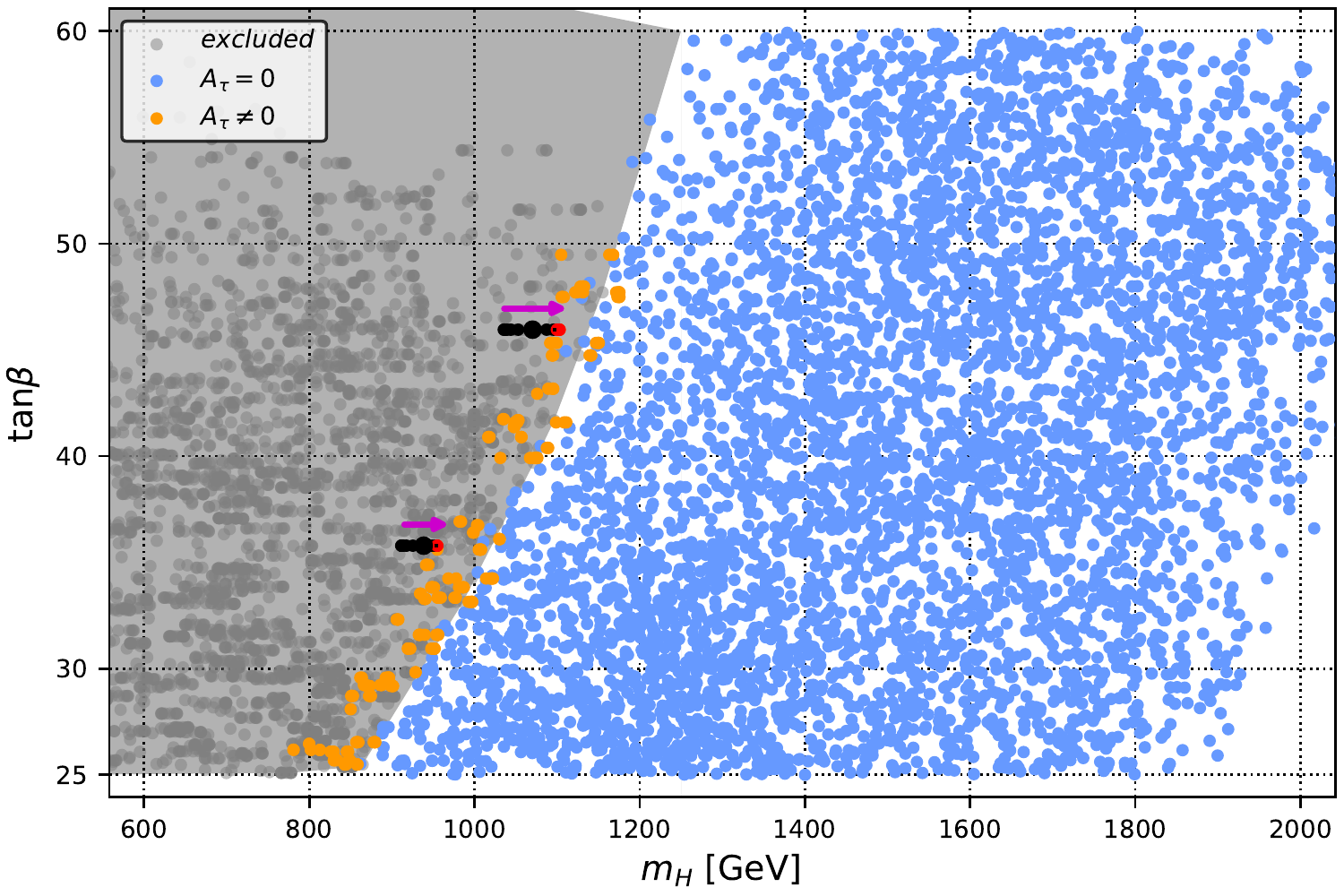}
 \caption{Tan $\beta$ vs  $m_H$ $[{\rm GeV}]$. Same colour coding for points as in Fig.~\ref{fig:Ab-BrTau-mH}. 
 }
 \label{fig:stau-mH-Tanb}
\end{figure}

The colour coding in Figs.~\ref{fig:Atau-BrTau-mH} to \ref{fig:stau-mH-Tanb}  is the same as for the sbottom case, with the obvious replacements. 

In Fig.~\ref{fig:Atau-BrTau-mH}  on the left, we see that we can still suppress
the Br$(H\to\tau\bar{\tau})$ via decays into stau pairs. However, in comparison
with the decays into sbottoms, the suppression is less effective which may be
related to the number of colour that enters in the sbottom decay case, as well
as the larger Yukawa coupling of the bottom-quark with respect to the
tau-quark, see Eq.~(\ref{Eq:sdowndecay}). Furthermore, the smallest values
of Br$(H\to\tau\bar{\tau})$ which are stable occur for somewhat small values of
$A_{\tau}$, $|A_{\tau}|\sim 600$ GeV. We have checked that larger values of
$A_{\tau}$ would lead to stronger suppressions of $Br(H\to\tau\bar{\tau})$,
however they are excluded by vacuum stability constraints. 

In Fig.~\ref{fig:Atau-BrTau-mH}  on the right, contrary to
the sbottom case, the dependence of $m_{H}$ on $A_{\tau}$ is much milder once
again due to the smaller Yukawa and the lack of colour for the stau case. Note
that the orange points for which $A_{\tau}\neq 0$ and  which are stable, start
at $m_{H}\gtrsim 800$ GeV. This can be understood by looking at
Fig.~\ref{fig:Atau-sigBrBB} on the left, where due to our scanning procedure
which starts with points that are barely ruled out by the $H\to\tau\bar{\tau}$
searches and considers $\tan\beta>25$, the lightest mass $m_H$ we can obtain
which is barely ruled out is $m_H\approx 800$ GeV. If we had chosen a lower
value of $\tan\beta$, we could have observed the stau effect for smaller values
of $m_H$. Nonetheless, by looking at the two example points and also at the
"width" of the orange region, we conclude that the effect of staus is much less
significant in allowing a larger parameter region than that of sbottoms. This
can also be seen in Fig.~\ref{fig:Atau-sigBrBB} on the right.

Finally in Fig.~\ref{fig:stau-mH-Tanb}, we translate the results to the
$m_H-\tan\beta$ plane. We see that indeed one can get most of the orange points
in the would-be excluded region, delimited by the grey envelope. Again we see
that the depth of the orange points in the grey envelope is much thinner
compared to the sbottom case of Fig.~\ref{fig:mH-Tanb}.

There has been an analytical study of the stau case in Ref.~\cite{Carena:2013iba}. It showed that, at the time, a suppression of up to 20 $\%$ of $\sigma_{bbH}\times \mathrm{Br}(H\to\tau\bar\tau)$ 
with respect to the case without SUSY decays was achievable for values of $A_{\tau}\sim 1.3$ TeV. We find slightly better results in our numerical study, as can been seen in Fig.~\ref{fig:Atau-sigBrBB} on the left,  where we estimate up to 50 $\%$ suppression for $\sigma_{bbH}\times \mathrm{Br}(H\to\tau\bar\tau)$ with respect to the case with no SUSY decays. Notice also that values of  $A_{\tau}\sim 1.3$ TeV are on the borderline of metastability, as shown in Fig.~\ref{fig:Atau-BrTau-mH} on the left and that though an analytical study for the stau case is consistent, a similar one for the sbottom case is not straightforward due to the large dependence of $m_H$ on $A_b$.


\section{Light Stops}
\label{sec:stops}

In the case of stops, given that we want only stops and not sbottoms to be light and that to obtain a mass for the lightest
Higgs $h$ of $m_h\approx 125$ GeV, which implies $A_t \sim m_{Q_3}\sim 2$ TeV,
we have in the end that one stop is light (mostly right-handed) while the other stop is
much heavier (mostly left-handed).  We also consider values of
$\tan\beta\in[25,60]$. As mentioned in section \ref{sec:theory}, the way to
increase the branching ratio of H into stops is by increasing the value of
$\mu$. However, there are large radiative corrections  to the heavy Higgs mass $m_H$ which are much stronger than in the case of
$A_b$ or $A_{\tau}$ for the sbottom and stau cases. Thus the scanning procedure
of leaving everything fixed except $\mu$ is much less efficient and we are only
able to retrieve stable points for $m_H> 2.6$ TeV and $\mu> 2.4$ TeV, with very small branching
ratio into stops. Vacuum stability is only an issue for the
very largest values of $\mu\gtrsim 4.5$ TeV. Performing a random scan we were
able to see the effect of stops reducing the Br$(H\to\tau\bar{\tau})$ via a
Br$(H\to\tilde{t}_1\tilde{t}_1^*)\lesssim 0.4$. Their effect seems
to start at $m_H\gtrsim 1$  TeV and extend up to $m_H\approx 2.2$ TeV for
values of $\mu\in[1.2,3.5]$ TeV. The problem however in this case by performing
a random scan is that we loose the guide from the two sets of example points which we
showed for the sbottom and stau cases, respectively. Thus we decided to only
comment briefly on this possibility in reducing Br$(H\to\tau\bar{\tau})$.


\section{Conclusion}
\label{sec:conclusions}

Searches for heavy Higgs bosons decaying to a pair of $\tau$ leptons severely
constrain the parameter space of the MSSM for large $\tan\beta$. We
demonstrate three possible ways how to alleviate the constraints by new decay
channels into third-generation sfermion pairs. For large $\tan\beta$, the coupling of the heavy Higgs to the sbottoms and staus proportional to $\tan\beta$ can be further enhanced by a large value of the trilinear couplings $A_b$ and $A_\tau$, respectively, while the coupling to stops has a $\tan\beta$ independent part which can be enhanced by the SUSY conserving Higgs mass $\mu$. The maximum size of the trilinear couplings $A_f$ and $\mu$, however, is constrained by the stability of the electroweak vacuum.

Our numerical scan shows that light sbottoms have the greatest potential to
alleviate the constraints from heavy Higgs searches. After imposing vacuum
stability, $|A_b|$ can take values up to 2 TeV and leads to a reduction of the
branching ratio Br($H\to\tau\bar\tau)$ by more than a factor two down to
Br$(H\to\tau\bar\tau)\lesssim 0.05$, which enlarges the available parameter
space. Similarly, for light staus we find values of $|A_\tau|\sim 1.3$ TeV with
a reduced branching ratio Br$(H\to\tau\bar\tau)\sim 0.07$, which allows to
slightly enlarge the allowed region of parameter space. Finally, light stops
allow very large values of $|\mu|$ close to 5 TeV. However, radiative
corrections to the heavy Higgs mass $m_H$ are large and substantially increase
it. It is still possible to observe the effect of a reduced
branching ratio Br$(H\to\tau\bar\tau)$ via an increased branching ratio for the
decay into light stops Br$(H\to\tilde t_1 \tilde t_1^*)\lesssim 0.4$, but
a detailed discussion would require to fix the heavy Higgs mass $m_H$ as much
as possible when increasing $|\mu|$.

Although future searches for heavy Higgs bosons decaying to $\tau$-pairs may tighten the constraints on the MSSM parameter space and eventually exclude the orange points in the figures, our main conclusion, that new decays to light third-generation fermions will alleviate the constraints from heavy Higgs search, holds irrespectively. This scenario can be tested by improving the reach of the searches for light third generation sfermions.


\section*{Acknowledgements}
We thank Carlos Wagner for useful discussions.  This work has been supported by the European Research Council (ERC) Advanced Grant Higgs@LHC.  This work was supported in part by the Australian Research Council. We acknowledge the use of \matplotlib\ and \julia. This research was supported by use of the Nectar Research Cloud, a collaborative Australian research platform supported by the National Collaborative Research Infrastructure Strategy (NCRIS).

\bibliography{draftbib}

\end{document}